\documentclass[aps,twocolumn,preprintnumbers,amsmath,amssymb,footinbib,prl]{revtex4}
\usepackage{graphicx}
\usepackage{bm}
\newcommand{\be}{\begin{equation}}
\newcommand{\ee}{\end{equation}}
\newcommand{\bea}{\begin{eqnarray}}
\newcommand{\eea}{\end{eqnarray}}

\newcommand{\lsim}{\mbox{\raisebox{-0.6ex}{$\stackrel{<}{\sim}$}}\:}

\begin{document}
\title{The role of local and global 
strangeness neutrality at the inhomogeneous freeze-out
in relativistic heavy ion collisions}
\author{ Detlef {\sc Zschiesche}\footnote{detlef@if.ufrj.br} and 
Licinio {\sc Portugal}\footnote{licinio@if.ufrj.br}}
\affiliation{Instituto de F\'\i sica, 
Universidade Federal do Rio de Janeiro\\
C.P. 68528, Rio de Janeiro, RJ 21941-972, Brazil}

\begin{abstract}
The decoupling surface in relativistic heavy-ion collisions
may not be homogeneous. Rather, 
inhomogeneities should form when a rapid
transition from high to low entropy density occurs. We analyze the
hadron "chemistry" from high-energy heavy-ion reactions for the
presence of such density inhomogeneities.
We show that due to the non-linear dependence of the particle
densities on the temperature and baryon-chemical potential such
inhomogeneities should be visible even in the integrated,
inclusive abundances. We analyze experimental data from
Pb+Pb collisions at CERN-SPS and Au+Au collisions at BNL-RHIC to
determine the amplitude of inhomogeneities and the role of 
local and global strangeness neutrality.
\end{abstract}

\maketitle

\section{Introduction}
It is expected that at sufficiently high
energies, a transient state of deconfined matter with broken $Z(3)$
center symmetry and/or with (approximately) restored chiral symmetry
is produced in collisions of heavy nuclei.
Lattice QCD simulations~\cite{Fodor:2001pe} indicate that a
second-order critical point exists, which was predicted by effective chiral
Lagrangians~\cite{SRS}; present estimates locate it at
$T\approx160$~MeV, $\mu_B\approx360$~MeV.  This point, where the
$\sigma$-field is massless, is commonly assumed to be the endpoint of
a line of first-order phase transitions in the $(\mu_B,T)$ plane.
To detect that endpoint, it is hoped 
that by varying the
beam energy, for example, one can ``switch'' between the regimes of
first-order phase transition and cross over, respectively.  If the
particles decouple shortly after the expansion trajectory crosses the
line of first order transitions one may expect a rather inhomogeneous
(energy-) density distribution on the freeze-out
surface~\cite{Bower:2001fq,Paech:2003fe} (similar, say, to the CMB
photon decoupling surface observed by WMAP~\cite{wmap}).  On the other
hand, if the low-temperature and high-temperature regimes are smoothly
connected, pressure gradients tend to wash out density
inhomogeneities. Similarly, in the absence of phase-transition induced
non-equilibrium effects, the predicted initial-state density
inhomogeneities~\cite{iniflucs,spherio} should be strongly damped.

Thus, we investigate the properties of an inhomogeneous fireball at
(chemical) decoupling. Note that if the scale of these inhomogeneities
is much smaller than the decoupling volume then they can not be
resolved individually, nor will they give rise to large event-by-event
fluctuations. Because of the nonlinear dependence of the hadron
densities on $T$ and $\mu_B$, they should nevertheless reflect in the
{\em average} abundances. 

Our basic assumption is that as the fireball expands and cools, at
some stage the abundances of hadrons ``freeze'', keeping memory of the
last instant of chemical equilibrium. This stage is refered to as
chemical freeze-out. By definition, only processes that conserve
particle number for each species individually, or decays
of unstable particles may occur later on.
The simplest model is to treat the gas of hadrons within the grand
canonical ensemble, assuming a homogeneous decoupling volume.  The
abundances are then determined by two parameters, the temperature $T$
and the baryonic chemical potential $\mu_B$; 
the chemical potential
for strangeness is fixed by the condition for overall strangeness
neutrality.
Fits of
hadronic ratios were performed extensively~\cite{Pbm05,thermo} 
within this
model, sometimes also including a strangeness ($\gamma_s$) 
or light quark ($\gamma_q$)  supression factor  
\cite{thermo_gammas,Becc05} or interactions with the chiral
condensate~\cite{thermo_int}.

In \cite{PRC} we analyzed the experimental data
on relative abundances of hadrons with respect to the presence of
inhomogeneities on the decoupling surface. 
To that end we proposed a
very simple and rather schematic extension of the common grand
canonical freeze-out model, i.e.\ a {\em superposition} of such
ensembles with different temperatures and baryon-chemical
potentials. Each ensemble is supposed to describe the
local freeze-out on the scale of the correlation length $\sim 1/T\sim
1-2$~fm. Even if freeze-out occurs near the critical point, the
correlation length of the chiral condensate is bound from above by
finite size and finite time effects, effectively resulting in similar
numbers~\cite{corr_length}. On the other hand, for small chemical
potential, far from the region where the $\sigma$-field is critical,
the relevant scale might be set by the correlation length for Polyakov
loops, which is of comparable
magnitude~\cite{loop_corlength}. Classical nucleation theory for
strong first-order phase transitions predicts even larger
``bubbles''~\cite{nucleation} but is unlikely to apply to small,
rapidly expanding systems encountered in heavy-ion
collisions~\cite{Paech:2003fe,Scavenius:2000bb}. Another (classical)
model for the formation of small droplets in rapidly expanding QCD
matter has been introduced in~\cite{Mishustin:1998eq}.
The entire decoupling surface contains many such ``domains'', even if
a cut on mid-rapidity is performed. We therefore
expect that the distributions of temperature and chemical potential
are approximately Gaussian \cite{future}. 
Besides simplicity, another goal of the
present analysis is to avoid reference to a particular dynamical
model for the formation or for the distribution of density
perturbations. In fact, we presently aim merely at checking whether
any statistically significant signal for the presence of
inhomogeneities is found in the data. If so, more sophisticated
dynamical models could be employed in the future to understand the
evolution of inhomogeneities from their possible formation in a phase
transition until decoupling. 

Rate equations for nuclear fusion and dissociation processes (and
neutron diffusion) have been used for inhomogeneous big bang
nucleosynthesis in the early universe~\cite{iBBN}. Similarly, hadronic
cascade models could be used for heavy-ion
reactions~\cite{HydroCascade}. This would remove reference to the
grand-canonical ensemble and to a thin decoupling {\em surface} in
space-time. In fact, hadronic binary rescattering models do predict a
rather thick freeze-out layer~\cite{HydroCascade,sorge}, where matter
expands non-ideally. On the other hand, the steep drop of
multi-particle collision rates with temperature should narrow the
freeze-out again~\cite{BSW}. In either case, we do not expect a strong
energy dependence of the width of freeze-out (see also~\cite{ceres}).

At chemical freeze-out, matter is in a state of expansion. However,
such flow effects do not affect the relative abundances of the particles
(in full phase space) if their densities are homogeneous throughout
the decoupling volume. The total number of particles of species $i$,
integrated over a solid angle of $4\pi$, is given by an integral of
the current $N_i^\mu = \rho_i\,u^\mu$, with $u^\mu$ the four-velocity of the
expanding fluid, over a given freeze-out hypersurface $\sigma^\mu =
(t^{\rm fo},\vec{x}^{\rm \, fo})$: 
\be N_i = \int d\sigma_\mu N_i^\mu =
\rho_i(T^{\rm fo},\mu_B^{\rm fo}) \int  u^\mu d\sigma_\mu~.  
\ee 
The second factor on the r.h.s.\ is nothing but the three-volume $V_3$
of the decoupling hypersurface as seen by the observer. This volume is
common to all species and drops out of multiplicity ratios: $N_i/N_j =
\rho_i^{\rm fo}/\rho_j^{\rm fo}$. It is clear that the argument holds
even when cuts in momentum space are performed, provided that the
differential distributions of all particles do not depend on that
particular momentum-space variable (for example, rapidity cuts for
boostinvariant expansion~\cite{CleyRed99}).

When the intensive variables $T$ and $\mu_B$ vary,
then the integration measure $(\int u\cdot d\sigma)/V_3$
will, in general, depend on the assumed distribution and amplitude of
inhomogeneities, as well as on the hydrodynamic flow profile etc.\
Nevertheless, it is still the same for all particle species and so can
be written in the form
\be
{1\over V_3}\int u\cdot d\sigma \longrightarrow \int dT d\mu_B \, P(T,\mu_B)~,
\ee
with $P(T,\mu_B)$ some distribution for $T$ and $\mu_B$.
For
simplicity, and for lack of an obvious motivation for assuming
otherwise, we shall take $P(T,\mu_B)$ to factorize into a distribution
for $T$, times one for $\mu_B$. These distributions could, in
principle, be obtained from the real-time evolution of the phase
transition~\cite{Bower:2001fq,Paech:2003fe}.

\section{The model}
\label{model}
In \cite{PRC} we introduced our model to 
analyze the available data from heavy-ion collisions at CERN-SPS and BNL-RHIC.
There the hadron abundances are determined by four parameters: the
arithmetic means of the temperatures and chemical potentials of all domains,
$\overline{T}$ and $\overline{\mu}_B$, and the widths of their
Gaussian distributions, $\delta T$ and $\delta \mu_B$.
Of course, the densities of strange particles depend also on the
strangeness-chemical potential $\mu_S$, which we determined in
\cite{PRC}  by imposing
local strangeness neutrality. That means, the 
strange chemical potential in each single domain was fixed 
by demanding zero net strangeness there.
However, the effect of independent fluctuations of
$\mu_S$ should also be looked at, in particular for
collisions at low and intermediate energies ($\surd s_{NN} \lsim
15$~GeV). This may help for example to reproduce the $\overline\Lambda$ to
$\overline{p}$ ratio, which was found to be larger
than one~\cite{inh_mus} and the $K^+/\pi^+$ enhancement around
$E_{\rm{Lab}}/A=30$ GeV~\cite{Na49_data}.  
Allowing for such independent fluctuations, 
the hadron abundances depend on six parameters: the
arithmetic means of the temperatures and chemical potentials of all domains,
$\overline{T}, \overline{\mu}_B$ and $\overline{\mu}_S$, 
and the widths of their
Gaussian distributions, $\delta T, \delta \mu_B$ and $\delta \mu_S$.
They read: 
\bea \label{ave_dens}
& &\overline{\rho}_i\; (\overline{T},\overline{\mu}_B, \overline{\mu}_S, \delta
T,\delta\mu_B, \delta\mu_S ) = 
\int\limits_0^\infty dT \;
P(T;\overline{T},\delta T) \\
& &\int\limits_{-\infty}^\infty d\mu_B \;
P(\mu_B; \overline{\mu}_B,\delta\mu_B) 
\int\limits_{-\infty}^\infty d\mu_S \; 
P(\mu_S; \overline{\mu}_S,\delta\mu_S)
~\rho_i (T,\mu_B, \mu_S)~, \nonumber
\eea
with $\rho_i(T,\mu_B,\mu_S)$ the actual ``local'' density of species
$i$, and with $P(x; \overline{x},\delta x) \sim \exp
[-{\left(x-\overline{x}\right)^2}/{2\, \delta x^2} ]$
the distribution of temperatures and chemical potentials within the
decoupling three-volume (the proportionality constants
normalize the distributions over the intervals where they are
defined). 
In addition, strangeness conservation enters now as a global
constraint for the mean of the 
strange chemical potential $\overline{\mu}_S$:
\bea \label{fs0eq}
f_s = \sum_i \overline{\rho}_i\; (\overline{T},\overline{\mu}_B, \overline{\mu}_S, \delta
T,\delta\mu_B, \delta\mu_S ) (n_s^i - n_{\overline{s}}^i) =0,
\eea 
with $f_s$ the net-strangeness, $n_s^i,n_{\overline{s}}^i$ the number of 
strange and anti-strange quarks of hadron species $i$, respectively. 
That means, the global densities obtained for given values of  
temperature and chemical potential parameters
weighted with the corresponding net number of strange quarks 
are summed and demanded to vanish to guarantee strangeness neutrality.
In the limit $\delta T$, $\delta\mu_B, \delta\mu_S \to0$
the Gaussian distributions are replaced by $\delta$-functions and the
conventional homogeneous freeze-out scenario is recovered:
\begin{equation}
\label{rhohom}
\overline{\rho}_i\; (\overline{T},\overline{\mu}_B, \overline{\mu}_S,0,0,0) =
\rho_i(\overline{T},\overline{\mu}_B,\overline{\mu}_S), 
\end{equation}
and the 
corresponding strangeness neutrality condition fixing $\overline{\mu}_S$.
In other words, in that limit
the average densities are uniquely determined by the first moments of
$T$ and $\mu_B$.
For the present investigation, 
we set the width of the distribution for the 
strange chemical potential equal to zero, $\delta\mu_S =0$.
Since eq.(\ref{fs0eq}) only ensures global strangeness neutrality, 
in this limit still  
finite net strangeness values in individual domains will appear, 
in contrast to our former analysis, where 
we fixed $\mu_S$ by $f_s=0$ locally.
It is important to note that 
with setting $\delta\mu_S =0$ and the global constraint eq. \ref{fs0eq} 
for $\overline{\mu_S}$, the densities again are a function
of four parameters: 
$\overline{T},\overline{\mu}_B,\delta T$ and  $\delta\mu_B$.
Thus we will write all quantities again as a function 
of these four parameters only.
In the following we will investigate how the 
fits to the experimentally measured particle abundances 
are influenced by the different strangeness neutraliy
conditions \cite{finitedmus}.

For the present analysis we compute the densities $\rho_i (T,\mu_B)$
in the ideal gas approximation, supplemented by an ``excluded volume''
correction:
\be \label{DensityExclVol}
\rho_i (T,\mu_B) = \frac{\rho_i^{\rm id-gas} (T,\mu_B)}{1+ 
v_i \sum_j \rho_j^{\rm id-gas} }~.
\ee
This schematic correction models repulsive interactions among the
hadrons at high densities. $v_i$ denotes the volume occupied by a
hadron of species $i$; we employ $v=\frac {4}{3} \pi {R_0}^3$ with
$R_0= 0.3$~fm for all species~\cite{ExclVol}. Therefore, for the homogeneous
model the denominator in~(\ref{DensityExclVol}) drops out of
multiplicity ratios. This is not the case for an inhomogeneous
decoupling surface, where the distributions of various species differ.
For all fits over the full solid angle, we
fixed the isospin chemical potential by equating the total charge in
the initial and final states; for the mid-rapidity fits at high
energies, we fixed $\mu_I=0$.

To illustrate the effect of inhomogeneities on the distributions of
various hadrons within the decoupling volume we introduce
\bea
& & D_i(T;\overline{T},\overline{\mu}_B, \delta T,\delta\mu_B)
      = P(T;\overline{T},\delta T) \nonumber \\ & &\hspace*{1cm}\times
\frac{\int\limits_{-\infty}^\infty d\mu_B \; P(\mu_B;
\overline{\mu}_B,\delta\mu_B)~\rho_i (T,\mu_B)~}
{\overline{\rho}_i\; (\overline{T},\overline{\mu}_B, \delta
  T,\delta\mu_B)}~, \label{D_T}  \\
& &\nonumber \\
& & D_i(\mu_B;\overline{T},\overline{\mu}_B, \delta T,\delta\mu_B)
      =P(\mu_B;\overline{\mu}_B,\delta\mu_B)
\nonumber \\ & &\hspace*{1cm}\times
\frac{\int\limits_0^\infty 
dT \;P(T;\overline{T},\delta T)~\rho_i (T,\mu_B)~}
{\overline{\rho}_i\; (\overline{T},\overline{\mu}_B, \delta
  T,\delta\mu_B)}~.\label{D_mu}
\eea
$D_i(T)$, for example, is the probability that a particle of type $i$ 
was emitted from a domain of temperature $T$. 
The main contribution to the integrals in~(\ref{ave_dens}) is {\em
  not} from $\overline{T}$ and $\overline{\mu}_B$ since hot spots
shine brighter than ``voids''. Rather, they are
dominated by the stationary points of the distributions
defined in eqs.~(\ref{D_T},\ref{D_mu}) above. Hence, the average
emission temperature $\langle
T\rangle_i$ and baryon-chemical potential $\langle\mu_B\rangle_i$
in general depend on the particle species $i$, unless $\delta
T=\delta\mu_B=0$. They can be evaluated as
\bea
\langle T\rangle_i &=& \int\limits_0^\infty 
    dT \;T\;  D_i(T;\overline{T},\overline{\mu}_B,
    \delta T,\delta\mu_B)~, \nonumber\\
\langle\mu_B\rangle_i &=& \int\limits_{-\infty}^\infty d\mu_B \;\mu_B\;
D_{i}(\mu_B;\overline{T},\overline{\mu}_B, \delta T,\delta\mu_B)~.
\label{averages}
\eea
Physically, this means that for non-zero widths of the temperature and
chemical potential distributions the freeze-out volume is not
perfectly ``stirred'', in that the relative concentrations of the
particles vary.

\section{Data analysis}

To determine the four parameters of the model we minimize
\begin{equation} \label{chi2}
\chi^2 = \sum_i {\left(r_i^{exp} - r_i^{model}\right)^2}/{\sigma_i^2}
\end{equation}
in the space of $\overline{T}$, $\overline{\mu}_B$, $\delta T$, and
$\delta\mu_B$. That is, we obtain least-square estimates for the
parameters, assuming that they are independent.
In~(\ref{chi2}), $r_i^{exp}$ and
$r_i^{model}$ denote the experimentally measured and the calculated
particle ratios, respectively, and $\sigma_i^2$ is set by the
uncertainty of the measurement.  Wherever available, we sum systematic
and statistical errors in quadrature.

The data used in our analysis are the particle multiplicities measured
by the NA49 collaboration for central Pb+Pb collisions at beam energy
$E_{\rm Lab}/A=20$, 30, 40, 80 and 158~GeV~\cite{Na49_data}, and those
measured by STAR for central Au+Au collisions at BNL-RHIC,
ref.~\cite{RHIC130_data} 
($\surd{s_{NN}}= 130$~GeV, compiled in \cite{RHIC_130_comp})
and ref.~\cite{RHIC_data} (200~GeV).
At RHIC energies, we analyze the midrapidity data; at top SPS energy,
both, midrapidity and $4\pi$ data.  At all other energies, we restrict
ourselves to the $4\pi$ solid angle data by NA49 in order to avoid
biases arising from differing acceptance windows of various
experiments. Furthermore, our checks showed that the fit results can
depend somewhat on the actual selection of experimental ratios. Hence,
where possible, we have opted for the least bias by choosing $r_i^{exp}
= N^{exp}_i / N^{exp}_\pi$, i.e.\ the multiplicity of species $i$
relative to that of pions. This represents the maximal set of
independent data points, as it is equivalent to fitting {\em absolute
multiplicities} with an additional overall three-volume parameter,
$N_i = V_3 \rho_i$. 

Specifically, at $E_{\rm Lab}/A=20$, 30, and 80~GeV the multiplicities
of $\pi^+$, $\pi^-$, $K^+$, $K^-$, $B-\overline{B}$, $\Lambda$,
$\overline\Lambda$, and $\phi$ are available. For the (in-)homogeneous
model, this leaves five (three) degrees of freedom. At 40~GeV, we can add
the $\Xi^-$ and $\Omega+\overline{\Omega}$.  The data sets for
top SPS energies include yet a few more species: $p$, $\overline{p}$ (only
midrapidity), $K^0_S$ (only $4\pi$), $\overline\Xi^+$ and $\Omega$,
$\overline\Omega$ seperately. For RHIC-130, we fitted to the
$K^+/K^-$, $\overline{p}/p$, $\overline{\Lambda}/\Lambda$,
$\Xi^+/\Xi^-$, $\overline{\Omega}/\Omega$, $K^-/\pi^-$, $K^0_S/\pi^-$,
$\overline{p}/\pi^-$, $\Lambda/\pi^-$, $K_0^{\ast}/\pi^-$,
$\phi/\pi^-$, $\Xi^-/\pi^-$ and $\Omega/\pi^-$ ratios.  Finally, at RHIC-200
the $K^+/K^-$, $\overline{p}/p$,
$\overline{\Omega}/\Omega$, $K^-/\pi^-$, $\overline{p}/\pi^-$,
$\Lambda/\pi^-$, $\overline{\Lambda}/\pi^-$, $\Xi^-/\pi^-$,
$\Xi^+/\pi^-$, $\Omega/\pi^-$, $\phi/K^-$ and $K_0^{\ast}/K^-$ ratios
were used. The first three ratios are close to unity and essentially
just set the chemical potentials to zero; they do not help to fix
$\overline{T}$, $\delta T$ and $\delta\mu_B$.

Where appropriate, feeding from strong and electromagnetic decays has
been included in $r^{model}_i$ by replacing ${\rho}_i
\rightarrow {\rho}_i + B_{ij}\; {\rho}_j.$ The
implicit sum over $j\neq i$ runs over all unstable hadron species,
with $B_{ij}$ the branching ratio for the decay $j\to i$, which were
taken from \cite{PDG}. From all the resonances listed by the Particle
Data Group \cite{PDG}, mesons up to a mass of 1.5~GeV and baryons up
to a mass of 2~GeV were included, respectively.  The finite widths of
the resonances were not taken into account, and unknown branching
ratios were excluded from the feeding.  These details are irrelevant
for the qualitative behavior of $\delta T$ and $\delta \mu_B$ but do,
of course, matter for quantitative results.

\section{Results}

\begin{figure}[h]
\vspace{-1.0cm}
\includegraphics[width=9cm]{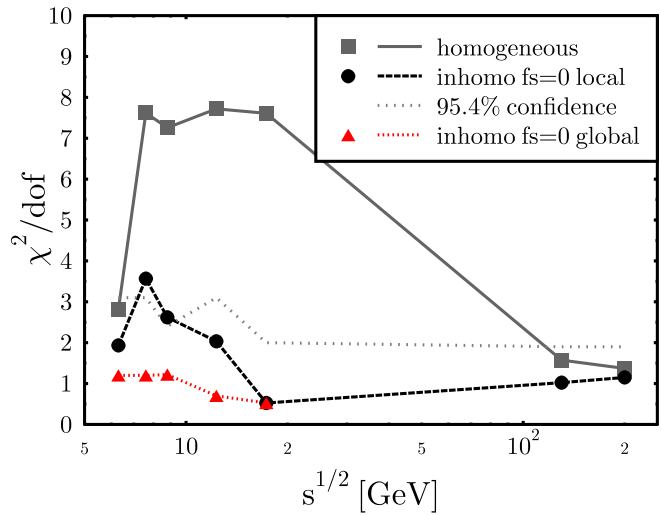}
\vspace{-0.2cm}
\caption{\label{chi2vssqrts}
$\chi^2/dof$ versus $\surd{s_{NN}}$ for the homogeneous
($\delta T = \delta \mu = 0$, squares) and the inhomogeneous fit
($\delta T$ and $\delta \mu$ free parameters, circles and triangles). Circles
denote the case of local strangeness neutrality, while
triangles represent the global strangeness neutrality case.
The lines are meant to guide the eye. Furthermore,
the $\chi^2/dof$ corresponding to the $95.4 \%$ confidence interval
is shown by the dotted line.}
\end{figure}
Fig.~\ref{chi2vssqrts} shows the minimal $\chi^2$ per degree
of freedom (taken as the number of data points minus the number of
parameters) for the homogeneous approach and the inhomogeneous approach
with local or global strangeness neutrality,
respectively. Note that the $\chi^2$-values for the homogeneous model 
are in general agreement with 
the analysis done in \cite{Pbm05} and other data from the literature 
\cite{thermo,thermo_gammas,RHIC_data}.
As already shown in \cite{PRC} and in general accordance with 
the analysis done in \cite{Pbm05}, 
for intermediate SPS energies, $E_{\rm Lab}/A
\simeq 30-160$~GeV, $\chi^2/dof$ 
is considerably smaller for the
inhomogeneous freeze-out surface than for the homogeneous case, which
is far outside the 95.4\% confidence interval \cite{spsfits}.
At $E_{\rm Lab}/A=20$~GeV and at RHIC energies,
$\chi^2/dof$ is similar for the inhomogeneous approach with local strangeness 
neutrality and the homogenous 
model. However, between 20 and 80 GeV the $\chi^2/dof$ values 
for the inhomogenous approach with local strangeness neutrality 
are rather large (between 2 and 4). 
In contrast, the inhomogenoues model 
with global strangeness neutrality gives  
$\chi^2/dof \approx 1$ for 
$E_{\rm Lab}/A \simeq 20-160$~GeV. It is important to note that 
this result is not due to introducing an additional parameter, but just 
due to allowing for domains of finite strangeness with global 
strangeness neutrality!
The calculations using global strangeness conservation for RHIC energies are under way, but due to the corresponding small baryon chemical potentials at these high energies no 
considerable effect should be expected.
Thus, the inhomogenoues model allowing for domains of finite net strangeness gives a very satisfactory  description 
($\chi^2/dof \approx 1$) of the experimental data for particle abundance ratios
from lowest SPS energies up to highest RHIC energies.
However, at RHIC the homogeneous approach already gives a good description of the data
and the inhomogeneous
model does not provide a statistically significant improvement. Thus,  
the assumption of a
nearly homogeneous decoupling surface can not be rejected 
there. On the other hand, the considerable improvement of the 
description of the data for 
$E_{\rm Lab}/A \simeq 30-160$~GeV 
indicates that at intermediate and high SPS energies, the experimental data 
favor an inhomogeneous freeze-out surface.
For the SPS 20 GeV data the situation is not clear: 
there is certainly a reduction of the $\chi^2/dof$ 
in the inhomogeneous approach, but also the 
homogeneous model gives a much better value than for higher SPS
energies. 
Here more experimental data are necessary to clarify the picture.
It is worth noting that in general, the
improvement due to the inhomogenoues decoupling surface 
is not driven by one single species; rather, the
inhomogeneous model describes nearly all multiplicities better than a
homogeneous decoupling surface~\cite{Bormio}. 

\begin{figure}[h]
\vspace{-1.0cm}
\includegraphics[width=9cm]{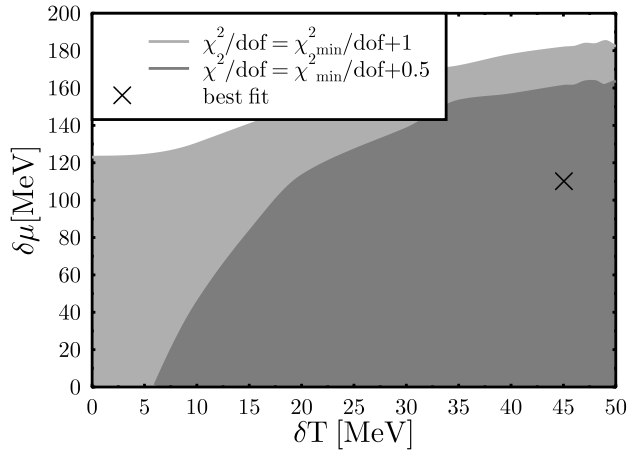}
\vspace{-0.2cm}
\caption{\label{dtdmu_rhic}
$\chi^2/dof$ contours in the $\delta T$, $\delta\mu_B$ plane for top
  RHIC energy, ($\surd{s_{NN}}=200$~GeV). 
The other two parameters
  ($\overline{T}$, $\overline{\mu}_B$) are allowed to vary freely. The
  $\chi^2/dof$ minimum is indicated by the cross.}
\end{figure}

\begin{figure}[h]
\vspace{-1.0cm}
\includegraphics[width=9cm]{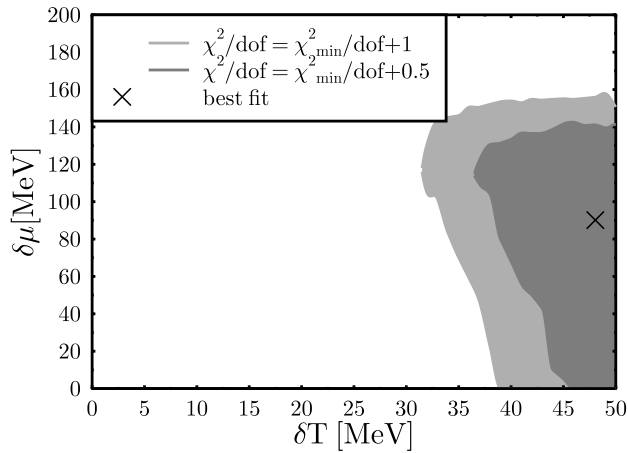}
\vspace{-0.2cm}
\caption{\label{dtdmu_sps158loc}
Same as fig.~\protect\ref{dtdmu_rhic} 
for top SPS energy
($E_{\rm Lab}=158$~GeV) with local strangeness neutrality.}
\end{figure}

\begin{figure}[h]
\vspace{-1.0cm}
\includegraphics[width=9cm]{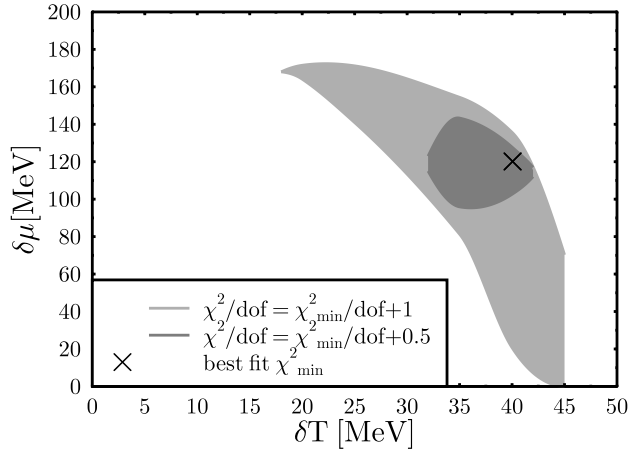}
\vspace{0.2cm}
\caption{\label{dtdmu_sps158glob}
Same as fig.~\protect\ref{dtdmu_rhic} 
for  top SPS energy
($E_{\rm Lab}=158$~GeV)
with global strangeness neutrality.}
\end{figure}

To illustrate the significance of inhomogeneities differently, we show
contours of $\chi^2/dof$ in the plane of $\delta T$, $\delta\mu_B$ in
figs.\ref{dtdmu_rhic}, \ref{dtdmu_sps158loc},  and ~\ref{dtdmu_sps158glob}. 
Here,
$\overline{T}$ and $\overline{\mu}_B$ were allowed to vary freely such as
to minimize $\chi^2$ at each point. 
Fig.~\ref{dtdmu_rhic} shows that at RHIC
energy, $\chi^2$ is very flat in both directions. This shows again that 
with the present
data points, a homogeneous freeze-out model appears to be a reasonable
approximation at high energies.
In contrast, fig.~\ref{dtdmu_sps158loc} shows 
that $\chi^2$ is relatively flat along the
$\delta\mu_B$ direction, while $\delta T$ is determined more
accurately and is clearly non-zero. In general we find that in the approach 
with local strangeness neutrality there is 
little correlation between $\delta T$ and $\delta\mu_B$ and that 
about the minimum, $\chi^2$ is rather flat in $\delta\mu_B$ direction
for all energies.
Finally, fig.~\ref{dtdmu_sps158loc} shows the contours at SPS 158 for the 
case of global strangeness neutrality. Now, the 
$\chi^2$ determines the $\delta\mu_B$ more accurately, 
favoring relatively large finite values.
For $\delta T$, again, values different from 
zero are strongly favored, which, however,  
turn out to be generally a little bit smaller than in the local $f_s=0$ case.  
The better description of the data and the better accuracy in 
determining the width of the 
$\mu_B$-distribution can be explained as follows:
If vanishing net strangeness is demanded in each single domain, 
in regions with high baryon chemical potential 
the strange chemical potential has to be small to
guarantee $f_s=0$.
Thus, the possible increased production of strange particles
in domains with high baryon chemical potential  
is restricted and results in the shown flatness of the $\chi^2$ distribution 
in $\delta \mu_B$ direction. 
In contrast, if the net strangeness vanishes globally, 
in domains of high chemical potential resulting from a large width 
$\delta \mu_B$, a large number of strange particles can be produced. 
Thus, the $\chi^2$ should be much more sensitive to the value of 
$\delta \mu_B$.

\begin{figure}
\vspace{1cm}
\includegraphics[width=9cm]{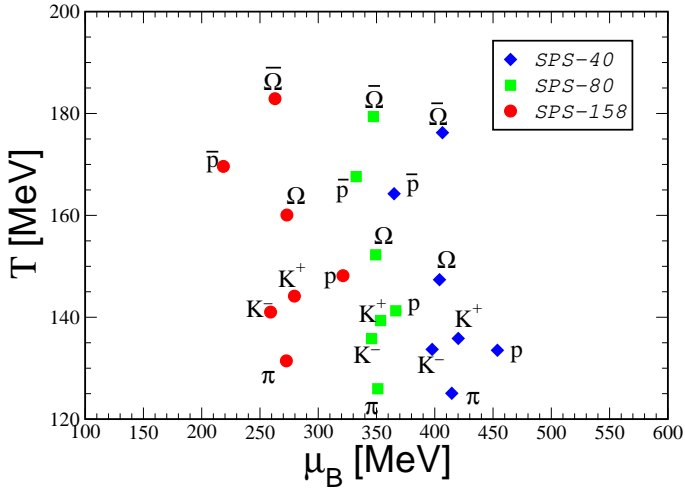}
\vspace{-0.2cm}
\caption{\label{foparticles}
Freeze-out temperatures $\langle T\rangle_i$
and chemical potentials $\langle \mu_B\rangle_i$ of various particle
species at $E_{\rm Lab}/A=40$, 80, 158~GeV for local strangeness
neutrality.}
\end{figure}

\begin{figure}
\vspace{1.0cm}
\includegraphics[width=9cm]{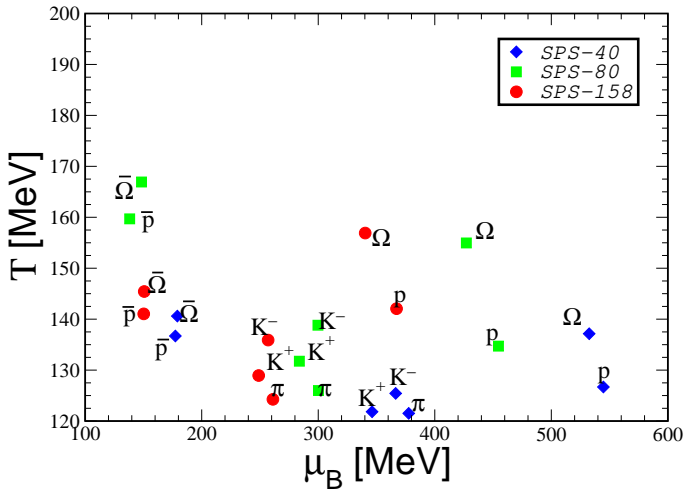}
\vspace{-0.2cm}
\caption{\label{foparticles_fsglob}
Freeze-out temperatures $\langle T\rangle_i$
and chemical potentials $\langle \mu_B\rangle_i$ of various particle
species at $E_{\rm Lab}/A=40$, 80, 158~GeV for global strangeness
neutrality.}
\end{figure}

As already discussed in section \ref{model}, an inhomogeneous freeze-out
surface or finite values for the width-parameters 
result in different mean emission temperatures 
and chemical potentials 
for different particle species, c.f. eq. \ref{averages}. 
These are shown in fig. \ref{foparticles} for the 
case of local strangeness neutrality and in fig. 
\ref{foparticles_fsglob} for the case of global strangeness 
neutrality at selected energies in the
CERN-SPS range. For the cases shown, the inhomogeneities determined from 
the fits to the particle abundances are large.
Note that the different values for these 
mean emission temperatures and chemical potentials 
result from the convolution of the distribution function for a given
particle species with the Gaussian probability distribution 
determined by the four parameters 
$\overline{T}, \overline{\mu_B}, \delta T, \delta \mu_B$.
For both cases, the effect of the inhomogeneities is evident.
For
example, anti-protons are typically emitted from regions with
lower baryon-chemical potential than protons; also, heavy particles
are concentrated in ``hot spots'' while light pions are
distributed more evenly throughout the decoupling volume etc.
\cite{hotspots}.
Figures \ref{foparticles} and \ref{foparticles_fsglob}
also show the differences in the resulting mean emission
temperatures and chemical potentials, depending on whether 
local or global strangeness neutrality is adopted:
In the case of local strangeness neutrality, the emission
chemical potentials of baryons and the corresponding anti-baryons do
differ much less than in the case of globally vanishing
net-strangeness. For example the mean emission 
baryon chemical potential for the $\Omega$ and the $\overline \Omega$
are nearly identical for $f_s = 0$ locally, while 
they are widely separated for the global constraint.
This results from the above discussed effect of the adjustement 
of the strange chemical 
potential in the local case. There,  
$\mu_S$ is so small (large negative value) in regions 
with large $\mu_B$, that the resulting 
chemical potential of $\Omega$ and $\overline \Omega$
are similar. This is not the case anymore if the 
strange chemical potential is determined globally and constant 
for the different domains. 
Then, the mean freeze-out points for the different particle 
species are spread over a much wider range in the 
$T-\mu_B$-plane.
On the other hand, the spread in temperature is somewhat larger in the 
local case than for global strangeness neutrality, resulting from the 
larger best fit values for the width parameters $\delta T$.
\section{Summary and Outlook}
In summary, we have shown that inhomogeneities on the freeze-out hypersurface do
not average out but reflect in the {\em 4$\pi$ (or midrapidity),
  single-inclusive} abundances of
various particle species. This is due to the non-linear dependence of
the hadron densities $\rho_i(T,\mu_B)$ on the local temperature and
baryon-chemical potential. Consequently, even the average
$\overline{\rho}_i$ probe higher moments of the $T$ and $\mu_B$
distributions. 
In \cite{PRC} we showed that an inhomogenoues freeze-out model 
with local strangeness conservation strongly improves the 
description of the data at medium and top SPS energies compared to 
the homogeneous freeze-out.
Here we showed that inducing global strangeness neutrality, 
results, without adding 
an additional parameter, 
in a further reduction of the $\chi^2$ at SPS energies.  
With the resulting $\chi/dof \approx 1$
for the whole range - from lowest SPS to 
highest RHIC energies.
Furthermore, while for local strangeness neutrality 
we observed a rather flat $\chi^2/dof$ in $\delta {\mu}_B$ direction,
this is determined more 
accuratrely if strangeness neutrality is ensured only globally. 
Rather in this approach a high statistical significance for 
a finite width of the distributions for temperature {\em and} 
baryon chemical potential at medium and high SPS energies is observed.
In addition we showed how in this region the mean emission 
temperature and chemical potential vary for different 
particle species. Our results also show that there are some characteristic
differences in the distribution of the resulting mean emission 
values, depending on whether strangeness neutrality is 
fulfilled locally or globally. Especially the 
separation in the mean emission chemical potential between baryons and
the corresponding anti-baryons is strongly influenced by the adopted 
strangeness neutrality condition.

Inhomogeneities could also affect the coalescence probabilities of
(anti-) nucleons to light (anti-) nuclei, which are also sensitive to
density perturbations~\cite{ioffe}. Other signals, such as two-particle
correlations~\cite{spherio,bubble}, could also be analyzed in this regard.
Future studies should
shed more light on whether these inhomogeneities can indeed be
interpreted as fingerprints of a first-order phase
transition. Eventually, one would want to establish more quantitative
relations between the amplitudes of the $T$, $\mu_B$ inhomogeneities
and the properties of the phase transition, e.g.\ its latent heat and
interface tension.  

Data from GSI-FAIR, the low energy program at RHIC and 
and CERN-LHC will provide additional constraints for the evolution of
chemical freeze-out with energy.

To improve the quality of the statistical fits, more data on hadron
multiplicities would be helpful, in particular at the lower end of the
CERN-SPS energy spectrum and at RHIC. This includes estimates of
multiplicities of unstable resonances ($\rho$, $K^*$, $\omega$,
$\Delta$ ...) at chemical freeze-out~\cite{reson}.  Data from GSI-FAIR
and CERN-LHC will provide additional constraints for the evolution of
chemical freeze-out with energy.

\vspace*{1cm} 
{\centerline{\em{\bf Acknowledgements}}} 
We thank A. Dumitru for cooperating in this work,
C.~Greiner for fruitful remarks concerning the model,
C.~Blume and M.~Gazdzicki for helpful discussions about the NA49 data and
A.~Grunfeld for helping with the construction of the resonance table,
CAPES and CNPq for partial support and 
the organizers of the workshop 
on 'Particle Correlations and Femtoscopy' 
September 9-11, 2006, Sao Paulo, Brazil,
where this work was presented.


\begin{thebibliography}{99}
\bibitem{future}
The investigation of 
other distributions and also the application of the ''superstatistics''
approach \cite{SuperStat} are under way.
\bibitem{inh_mus} C.~H\"ohne (for the NA49 collaboration), nucl-ex/0510049.
Our present best fit (of the $4\pi$ ratios) for the inhomogeneous
model without independent fluctuations of $\mu_S$ yields
$\overline\Lambda/\overline{p}=0.93$ at $E_{\rm Lab}/A=40$~GeV, for
example, versus 0.76 for $\delta T=\delta\mu_B=0$ (without
contributions from weak decays).
\bibitem{finitedmus}
Calculations for finite values of $\delta\mu_S =0$ are under way.
However, checking the influence of global strangeness conservation without 
adding an additional parameter represents the first step to be taken 
in this direction.
\bibitem{spsfits}
For SPS-158 only the
$4\pi$ fit is shown; restricting the homogeneous fit to the
mid-rapidity data gives smaller $\chi^2/dof$, but still significantly
higher than in the inhomogeneous approach. 
$\chi^2$ is smaller if other particle ratios are considered, as 
for example $\Xi/\Lambda, \Omega/\Xi$ instead of $\Xi/\pi, \Omega/\pi$
\cite{Pbm05} or if finite widths of resonances are taken into 
account \cite{Becc05}. However, the increase of $\chi^2$ 
at SPS energies is generic if Na49 $4\pi$-data are fitted.
\bibitem{hotspots}
 Note that the rather high temperatures of the hot spots
from which the heavy particles emerge might indicate the need for a better
treatment of interactions~\cite{thermo_int} than the simple
excluded-volume model employed here.
\end{thebibliography}
\end{document}